\documentclass[traditabstract]{aa} 
\usepackage{natbib} 
\usepackage{graphicx}
\usepackage{txfonts}
\usepackage{epsfig}
\usepackage[T1]{fontenc}
\usepackage[utf8]{inputenc}
\usepackage[english]{babel}
\usepackage{amsmath}
\usepackage{amssymb}
\usepackage{color}
\usepackage{comment}
\usepackage{wrapfig}
\usepackage{float}
\usepackage{subfig}
\usepackage{threeparttable}
\usepackage{empheq}
\usepackage{bm} 
\usepackage{listings}
\usepackage[toc,page]{appendix}
\usepackage{gensymb} 
\usepackage{booktabs}
\usepackage{caption}
\usepackage{cancel}

\begin{document}

   \title{The magnetospheric radius of an inclined rotator in the magnetically threaded disk model}

   \author{E. Bozzo 
          \inst{1}
          S. Ascenzi
          \inst{2,3,4}
          L. Ducci 
          \inst{5,1}
          A. Papitto
           \inst{2}
          L. Burderi 
           \inst{6}
           L. Stella 
           \inst{2}                               
          }
   \institute{Department of Astronomy, University of Geneva, Chemin d’Ecogia 16,
             CH-1290 Versoix, Switzerland; \email{enrico.bozzo@unige.ch}
             \and 
             INAF - Osservatorio Astronomico di Roma, Via Frascati 33, 00044 Rome, Italy  
             \and
              Universit\`a di Roma Tor Vergata, Via della Ricerca Scientifica 1, I-00133 Roma, Italy
             \and 
             Dip. di Fisica, Universita` di Roma La Sapienza, P.le A. Moro, 2, I-00185 Rome, Italy
             \and 
             Institut f\"ur Astronomie und Astrophysik, Kepler Center for Astro and Particle Physics, Eberhard Karls Universit\"at, Sand 1, 72076 T\"ubingen, Germany
             \and 
             Dipartimento di Fisica, Universit\'a degli Studi di Cagliari, SP Monserrato-Sestu km 0.7, 09042 Monserrato, Italy
             }
   
   \authorrunning{E. Bozzo et al.}
   \date{Submitted: -; Accepted -}

  \abstract{The estimate of the magnetospheric radius in a disk-fed neutron star X-ray binary 
  is a long standing problem in high energy Astrophysics. We review the magnetospheric radius calculations in the so-called 
  magnetically threaded disk model, comparing the simplified approach originally proposed by \citet{ghosh79b} with the revised version proposed by \citet{wang87},  
  \citet{wang95}, and \citet{wang97}. We show that for a given set of fixed parameters (assuming also a comparable screening factor of the neutron star magnetic 
  field by the currents induced on the disk surface) the revised magnetically threaded disk model predicts a 
  magnetospheric radius that is significantly smaller than that derived from the \citet{ghosh79b} treatment. For a fixed value of the 
  neutron star magnetic field and a wide range of mass accretion rates, the inclusion of a large inclination angle between the neutron 
  star rotation and magnetic field axes ($\chi$$\gtrsim$60~deg) leads to a further decrease of the magnetospheric radius.  
  To illustrate the relevance of these calculations, we consider, as an example, the case of the transitional pulsars. 
  During the so-called ``high mode'' of their sub-luminous accretion disk state, these sources have shown X-ray pulsations 
  interpreted as due to accretion at an unprecedented 
  low luminosity level compared to other neutron stars in X-ray binaries. In the context of the magnetic threaded disk model, 
  we show that accretion at luminosities of $\sim$10$^{33}$~erg~s$^{-1}$ (and thus accretion-driven 
  X-ray pulsations) can be more easily explained when the prescription of the magnetospheric radius provided by \citet{wang97} is used. This avoids   
  the need of invoking very strong propeller outflows in the transitional pulsars, as proposed in other literature works.}


   \maketitle

\section{Introduction}
\label{sec:intro}

In disk-fed X-ray binaries hosting neutron stars (NS), the determination of the so-called magnetospheric 
radius is a long-standing problem that has been faced through different theoretical approaches and numerical simulations 
\citep[see, e.g.][for a recent review]{lai14}. Different models aimed at determining self-consistently the magnetospheric radius have been challenged  
by our relatively poor knowledge of parameters related to the micro-physics of the disk matter (magnetic diffusivity, turbulence, ...), 
as well as its complex coupling with the NS magnetic field \citep[see, e.g.,][]{frank02}. 
Numerical simulations have highlighted some aspects of the disk-magnetosphere interaction \citep[see, e.g.][]{romanova14, parfrey, parfrey1}.  
However, from these simulations it is often difficult to derive simple prescriptions to be used in the interpretation 
of X-ray data of accreting NSs in X-ray binaries over a wide range of luminosity \citep[10$^{33}$-10$^{38}$~erg~s$^{-1}$; see, e.g.][]{darias14, walter15}.  
The position of the magnetospheric radius is, indeed, used as a proxy to predict if accretion can take place in these systems and  
accretion powered X-ray pulsations should be expected as a consequence of the channelling of the accreted material toward the magnetic poles of the 
compact object \citep[see, e.g.,][]{ba91,ghosh07,patruno12}.

Even though many different approaches have been proposed to estimate the magnetospheric radius \citep[see, e.g.,][and references therein]{klu,carol}, 
we limit in this paper to one of the most frequently-used analytical approximation that is available within the so-called ``magnetically threaded 
disk model'' as originally proposed by \citet[][hereafter GL79]{ghosh79b} and later revised by \citet[][hereafter WG87]{wang87} 
and \citet[][hereafter WG95]{wang95}. The basic assumptions concerning the coupling between the NS magnetic field and the 
disk are similar in GL79, WG87, and WG95, but the dependence of the magnetospheric radius on the different parameters is significantly 
different, as also discussed previously in \citet[][hereafter B09]{bozzo09}. 

In this work, we focus on the WG87 and W95 method to derive the magnetospheric radius, including the extension to the 
case of an oblique rotator, as presented later by \citet[][hereafter WG97]{wang97}. For comparable values of a  
number of the threaded disk model parameters (including the screening factor of the neutron star magnetic field by the currents induced on the 
disk surface), we highlight that the magnetospheric radius predicted by WG97 is significantly smaller than that expected from the 
original GL79 treatment. The reduction of the magnetospheric radius is more pronounced toward low mass accretion rates 
and for higher inclination angles between the NS magnetic and rotational axis (the effect of the high inclination angle 
is more effective toward higher mass accretion rates). 

These findings are applied, as an example, to the case of the so-called transitional millisecond pulsars, which display  
coherent X-ray pulsations interpreted as due to accretion at luminosities that are  $\sim$100 times lower than those of other NS X-ray binaries.

\section{The magnetospheric radius in the magnetic threaded disk model}
\label{sec:prescriptions}

\subsection{The aligned rotator case}
\label{sec:aligned}

Let us consider the case of a disk-fed X-ray binary in which the NS is an aligned rotator (i.e. with  
aligned magnetic and spin axes, which are both perpendicular to the plane of the disk). 
If the disk is not completely diamagnetic and the accreting plasma has a non-zero resistivity, then the NS magnetic field lines 
can penetrate inside the disk (the so-called ``magnetic threaded disk model''). These magnetic field lines 
regulate the accretion process and the exchange of angular momentum between the NS and the disk.
The momentum exchanged through the magnetic field lines penetrating the disk at radii smaller than the corotation 
radius\footnote{$R_{\rm c}$ corresponds to the distance from the NS 
at which the Keplerian angular velocity of the material in the disk ($\Omega_{\rm K}$) is equal to the NS angular velocity 
($\Omega_{\rm NS}$).}  
\begin{equation}
R_{\rm c} = 1.7\times10^6 M_{1.4} P_{-3}^{2/3}~{\rm cm} 
\end{equation}
contributes to spin up the NS, whereas that exchanged through lines penetrating the disk beyond $R_{\rm c}$ acts to reduce the 
star spin. Note that we scaled the NS mass, $M$, in units of 1.4~$M_{\odot}$ and its spin period, $P_{\rm spin}$, 
in units of 1~ms. Assuming that the NS has a dipolar magnetic field, we can write the z-component of the field close to the disk 
surface:   
\begin{equation}
B_{\rm z}(R)=-\eta\frac{\mu}{R^3}, 
\label{eq:bz} 
\end{equation}
where $\mu$ is the NS magnetic moment, $\eta\sim$0.2 is a screening 
parameter representing the effect of currents induced on the disk surface, and 
$R$ is the distance from the NS center. A toroidal magnetic field $B_{\phi}$ is 
generated from $B_{\rm z}$ due to the differential rotation between the star and the disk.  The shear amplification of the toroidal field 
occurs on a time scale $\tau\sim\vert\gamma(\Omega_{\rm NS}-\Omega_{\rm K})\vert^{-1}$, where $\gamma\sim1$ is a parameter describing 
the steepness of the vertical transition across the disk height, $h$, between the rigid corotation of the magnetic field line 
with the star and the Keplerian rotation inside the disk. Due to the finite conductivity of the disk material, the magnetic field  
lines distorted beyond a certain degree can reconnect above and below the plane of the disk on a time scale 
$\tau_{\rm phi}\sim h/(\xi v_{\rm A\phi})$, where $\xi v_{\rm A\phi}$ is the reconnection rate expressed 
in terms of the local Alfven velocity. The numerical factor $\xi$ is expected to be $\xi\simeq0.01-0.1$ if the main dissipation is the annihilation 
of the poloidal field near the disk midplane, or $\xi\simeq1$ if magnetic buoyancy is dominant. 

GL79 intuitively proposed that the amplification of the toroidal field could be described by: 
\begin{equation}
\frac{B_{\phi}}{B_{\rm z}} \simeq \mp \frac{\gamma(\Omega_{\rm NS}-\Omega_{\rm K})h}{\xi v_{\rm Az}},
\label{eq:toroidal} 
\end{equation}  
where the upper sign corresponds to the case $z>0$. 
An issue with the above definition, as spotted by WG87, is that the magnetic pressure generated 
by the wound field ($B_{\phi}/8\pi$) would exceed the thermal pressure $p$ in the disk at radii $>R_{\rm c}$, thus invalidating 
all calculations for the magnetic threaded disks beyond the corotation radius. 
Furthermore, the quantity $\int^{R_{\rm s}}_{R_{\rm M}}B_{\phi}B_{\rm z} R^2 dR$, where $R_{\rm M}$ is the 
magnetospheric radius and $R_{\rm s}$ is a screening radius beyond which the magnetic threading of the disk becomes negligible, 
would diverge for thin disks at large radii \citep[$R_{\rm s}\rightarrow\infty$;][hereafter S73]{shakura73}. Therefore, 
GL79 introduced the screening radius to artificially limit the integration and avoid divergence. 
The integrated quantity mentioned above is important in the threaded disk model, as it regulates the torque acting on the NS 
that is produced by the star magnetic field lines penetrating the accretion disk. The magnetospheric radius can be calculated by equating  
the rate at which the NS magnetic field removes angular momentum from the disk and viscosity transfers it at larger radii in the disk:  
\begin{equation}
\frac{B_{\phi}(R_{\rm M})B_{\rm z}(R_{\rm M})}{\dot{M}(GMR_{\rm M})^{1/2}}=-\frac{1}{2(R_{\rm M})^3} 
\label{eq:rmdef} 
\end{equation}
(here $\dot{M}$ is the mass accretion rate). It is worth noting that Eq.~\ref{eq:rmdef} is only valid for $R_{\rm M}<R_{\rm c}$, 
as in the opposite case the magnetic field would add angular momentum to the disk rather than removing it. 
GL79 solved the above equations by assuming a boundary layer solution, in which material from the disk is progressively brought 
into corotation with the star by the magnetic field lines penetrating the disk. 
The boundary layer is divided in a broad outer zone extended between $R_{\rm M}$ and $R_{\rm s}$, where most of the threading takes place, 
and an inner boundary located within $R_{\rm M}$ and characterized by a limited radial extent ($\delta$R$\ll$$R_{\rm M}$). In the inner 
boundary, magnetic stresses are larger than viscous stresses and thus matter is lifted from the disk before being accreted onto the NS. 
Following this treatment, GL79 found that: 
\begin{eqnarray}
R_{\rm M}^{\rm GL79}\simeq 0.52 R_{\rm A} = 0.52 \mu^{4/7} (2GM)^{-1/7} \dot{M}^{-2/7} \nonumber \\
\simeq 1.6\times10^6 \mu_{26}^{4/7} M_{1.4}^{-1/7} \dot{M}_{16}^{-2/7}~{\rm cm}, 
\end{eqnarray}
where $\mu_{26}=\mu/10^{26}$~G~cm$^{3}$, $\dot{M}_{16}=\dot{M}/10^{16}$~g~$s^{-1}$, and $R_{\rm A}$ is the so-called Alfv\'en radius.  
$R_{\rm A}$ is usually considered a good approximation of the magnetospheric radius in case of spherical (as opposed to disk) 
accretion \citep[see discussion in B09 and][]{bozzo08}. In the following, we will use for convenience the non-dimensional quantity: 
\begin{equation}
x_{\rm GL79} = R_{\rm M}^{\rm GL79}/R_{\rm c}\simeq 0.94 \mu_{26}^{4/7} M_{1.4}^{-8/7} \dot{M}_{16}^{-2/7} P_{-3}^{-2/3}.   
\label{eq:rmGL} 
\end{equation}
In the GL79 approach, the magnetospheric radius is a good approximation of the inner disk 
radius, as the extension of the transition region between the disk and the closed NS magnetosphere ($\delta$R) is estimated to be only a few \% 
of the magnetospheric radius. We note, however, that in other approaches proposed for the calculation of the magnetospheric radius 
the properties of the transition region can be significantly different \citep[see, e.g.][]{erkut04}.    

WG87 derived a different version of Eq.~\ref{eq:toroidal} starting from the Faraday induction law and obtained: 
\begin{equation}
\frac{B_{\phi}}{B_{\rm z}} \simeq \pm \vert \frac{\gamma(\Omega_{\rm NS}-\Omega_{\rm K})h}{\xi v_{\rm Az}} \vert^{1/2}. 
\label{eq:toroidal2} 
\end{equation} 
This formulation solves the divergence as well as the magnetic pressure issues affecting the GL79 
treatment. In this context, the magnetospheric radius can be readily calculated from Eqs.~\ref{eq:rmdef} and 
\ref{eq:toroidal2}. 
If the disk is considered to be well approximated by the solution of S73, 
then $h$=$c_{\rm s}$/$\Omega_{\rm K}$ and $c_{\rm s}$=($p$/$\rho$)$^{1/2}$. Here, $c_{\rm s}$ is the sound velocity 
in the disk, $\rho$ its density, and $p$ the thermal pressure. For the innermost gas-pressure dominated region of the 
S73 disk (region B), the magnetospheric radius in units of the corotation radius is 
(see Eq.~16 in B09): 
\begin{eqnarray}
x_{\rm w}^{-211/80}\sqrt{1-x_{\rm w}^{3/2}}& = & \nonumber 
 1.4\times10^{-2}\sqrt{\xi\gamma^{-1}\eta^{-3}}
\alpha^{9/40}\cdot\\ 
& &\cdot\mu_{26}^{-3/2}
\mu_p^{1/4}m^{7/6} P_{\rm -3}^{211/120}\dot{M}_{16}^{4/5}.  
\label{eq:xalfb}
\end{eqnarray}
For the outer gas-pressure-dominated region of the disk (the C region), it is found (see Eq.~17 in B09): 
\begin{eqnarray}
x_{\rm w}^{-85/32}\sqrt{1-x_{\rm w}^{3/2}}&=& \nonumber 
 1.6\times10^{-2}\sqrt{\xi\gamma^{-1}\eta^{-3}}
\alpha^{9/40}\cdot\\ 
&& \cdot\mu_{26}^{-3/2}
\mu_p^{1/4}m^{7/6} P_{\rm -3}^{85/48}\dot{M}_{16}^{63/80}, 
\label{eq:xalfc}
\end{eqnarray}
For the region A of the S73 disk (where the disk is dominated by radiation pressure), 
we obtain:   
\begin{eqnarray}
x_{\rm w}^{-19/8}\sqrt{1-x_{\rm w}^{3/2}}& = & \nonumber 
 1.8\times10^{-2}\sqrt{\xi\gamma^{-1}\eta^{-3}}
\alpha^{1/4}\cdot\\ 
& &\cdot\mu_{26}^{-3/2}
m^{2/3} P_{\rm -3}^{19/12}\dot{M}_{16}.  
\label{eq:xalfa}
\end{eqnarray}
Note that the case of region A (not reported previously by B09) is added here for completeness. 
In all the above equations $\mu_{\rm p}$ is the mean molecular weight ($\mu_{\rm p}$=0.615 for matter characterized by solar metallicity) 
and $\alpha$ is the viscosity parameter. 
For transient X-ray binaries in outbursts, the latter is believed to span the range $\sim$0.1-0.4 
\citep{king12,king13,lii14}. Equations~\ref{eq:xalfb}, \ref{eq:xalfc}, and \ref{eq:xalfa} must be solved numerically 
to compute $x_{\rm w}$ as a function of all other parameters. The transition between region B and  
region A of the disk occurs when the magnetospheric radius is (S73)  
\begin{equation}
r_{\rm ab} = 2.3\times10^6 \alpha^{2/21} m^{7/21} \dot{M}_{16}^{16/21}~{\rm cm},  
\label{eq:rab}
\end{equation}
while the radius for the transition between region C and B is 
\begin{equation}
r_{\rm bc} = 2.7\times10^8 m^{1/10} \dot{M}_{16}^{2/3}~{\rm cm}.   
\label{eq:bc}
\end{equation}
By using the above equations, it can be seen that for a NS with a spin period as short as a few milliseconds, the magnetospheric 
radius is located inside the region A for mass accretion rates $\gtrsim$10$^{16}$~g~s$^{-1}$. The predicted height of the disk around this radius 
would be a factor of $\sim$10 larger than that allowed by a S73 disk ($h$/$R$$\sim$0.01), due to the steep dependence of the disk height 
in region A on the mass accretion rate. As all the equations being used here are strictly valid only in case of thin disks and we 
are mainly interested in the low mass accretion regime (see Sect.~\ref{sec:transitional}), we limit 
all our analyses to $\dot{M}_{16}$$\lesssim$1. 

\citet{wang95} investigated also the impact of slightly different prescriptions for the growth of the toroidal 
magnetic field, beside the one presented in Eq.~\ref{eq:toroidal2}. In particular, he considered the case in which 
the amplification of the toroidal field is limited by either diffusive decay due to turbulent mixing within the disk or 
by reconnections occurring within the NS magnetosphere. In these cases: 
\begin{equation}
\frac{B_{\phi}}{B_z}\simeq\frac{\gamma(\Omega_{\rm NS}-\Omega_{\rm K})}{\alpha\Omega_{\rm K}}, 
\label{eq:alf3} 
\end{equation}
or 
\begin{equation}
\frac{B_{\phi}}{B_z}=
\left\{
\begin{array}{lr}
\gamma_{\rm max}(\Omega_{\rm NS}-\Omega_{\rm K})/\Omega_{\rm K}, 
&\Omega_{\rm K}\gtrsim\Omega_{\rm NS};\\
\gamma_{\rm max}(\Omega_{\rm NS}-\Omega_{\rm K})/\Omega_{\rm NS},  
&\Omega_{\rm K}\lesssim\Omega_{\rm NS},  
\end{array}
\right.
\label{eq:alf4} 
\end{equation}
respectively ($\gamma_{\rm max}$ is a parameter representing the maximum 
value of the magnetic azimuthal pitch). By using the same procedure as before, one finds for the magnetospheric radius (B09): 
\begin{eqnarray}
x_{\rm w}^{-7/2}-x_{\rm w}^{-2}& = & 
2.38\times10^{-2}\alpha\gamma^{-1}\eta^{-2}\mu_{26}^{-2}\cdot \nonumber \\ 
&& \cdot m^{5/3} P_{\rm -3}^{7/3}\dot{M}_{16} 
\label{eq:xdiff} 
\end{eqnarray}
and
\begin{eqnarray}
x_{\rm w}^{-7/2}-x_{\rm w}^{-2}& = & 
2.38\times10^{-2}\gamma_{\rm max}^{-1}\eta^{-2}\mu_{26}^{-2}\cdot \nonumber \\
&& \cdot m^{5/3} P_{\rm -3}^{7/3}\dot{M}_{16}, 
\label{eq:xrec} 
\end{eqnarray}
respectively for Eq.~\ref{eq:alf3} and \ref{eq:alf4}. Note that these two prescriptions hold independently of the disk region where 
the magnetospheric radius is located. 

\begin{figure}
\centering
\includegraphics[scale=0.5]{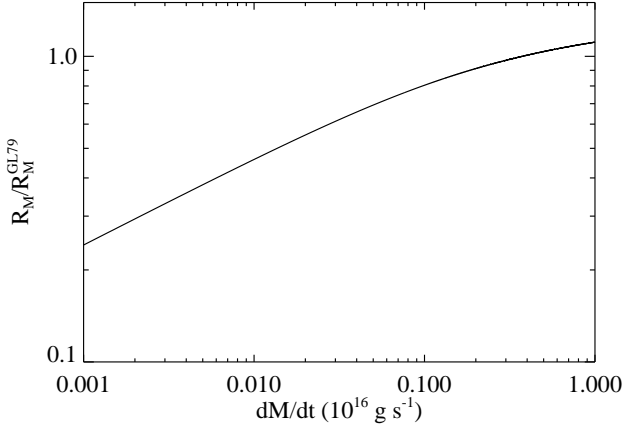}
\caption{Ratio between the GL79 magnetospheric radius and the magnetospheric radius calculated in the revised threaded disk model according to 
Eq.~\ref{eq:xdiff} (WG95) as a function of the mass accretion rate. We assumed $P_{\rm spin}$=1.69~ms, 
$\alpha$=0.4, $\eta$=0.2, $\mu_{26}$=0.78 (see Sect.~\ref{sec:inclined}), and $\gamma$=1.0.}     
\label{fig:plots1} 
\end{figure}

As discussed in B09, the magnetospheric radius estimated from  Eq.~\ref{eq:xalfb}, \ref{eq:xalfc}, 
\ref{eq:xalfa}, \ref{eq:xdiff}, and \ref{eq:xrec} is smaller than that computed from Eq.~\ref{eq:rmGL}, at 
low mass accretion rates (assuming a consistent value of $\eta$ between the different treatments). 
This is shown in Fig.~\ref{fig:plots1}, where we plot as a representative example the ratio between the 
magnetospheric radius calculated with Eq.~\ref{eq:rmGL} and Eq.~\ref{eq:xdiff} (as shown in B09, Eq.~~\ref{eq:xalfb}, \ref{eq:xalfc}, 
\ref{eq:xalfa}, \ref{eq:xdiff}, and \ref{eq:xrec} provide relatively similar results were compared to those of GL79). 
Note that we used in this figure $\eta$=0.2 for both Eq.~\ref{eq:rmGL} and Eq.~\ref{eq:xdiff}. 
The value $\eta$=0.2 was first suggested by GL79 solving in details the structure of the transition region between the closed NS 
magnetosphere and the accretion disk. The same value was adopted by WG87 in his revised version of the magnetically threaded disk model. A revised 
value of this parameter (as large as $\eta\simeq$1.0) was suggested by WG96 using the observations of quasi-periodic oscillations in X-ray pulsars 
and assuming these could be interpreted with the so-called beat frequency model \citep[BFM;][]{alpar85, lamb85}. 
The observational data were compared in WG96 with an approximate solution to the Eq.~\ref{eq:xdiff} obtained by assuming that the magnetic pitch 
$B_{\rm \phi}$/$B_{\rm z}$ of Eq.~\ref{eq:toroidal2} is constant and not depending from the radius. This approach was later revised by B09,  
who showed that when all parameter dependences are retained and more updated observations of X-ray pulsars are used, the application of the BFM is 
not straightforward and it is not possible to firmly conclude on the correct value of $\eta$ to be used. For this reason, we assume for the purpose of all  
analyses in this paper $\eta\simeq$0.2. This also allows us to carry out a self-consistent comparison 
between the magnetospheric radius originally derived by GL79 and the one revised by WG87, WG95, and WG97.  

\subsection{The inclined rotator case}
\label{sec:inclined}

The extension of the WG87 calculation of the magnetospheric radius to the case of an oblique rotator (i.e. when the NS magnetic and spin axes are not aligned) 
was presented by WG97. We summarize here his treatment\footnote{A simplified treatment of the magnetospheric radius in case of an oblique rotator 
was also presented by \citet{jetzer}.} and report the main equations that are needed for our scope. 
The NS magnetic field components in the radial, azimuthal, and vertical 
directions for an oblique rotator in the vicinity of the disk are:
\begin{align}
B_R&=2\eta\Bigl(\frac{\mu}{R^3}\Bigr)\sin\chi\cos\phi\\
B_\phi&=\eta\Bigl(\frac{\mu}{R^3}\Bigr)\sin\chi\sin\phi+b_\phi\notag\\
B_z&=\eta\Bigl(\frac{\mu}{R^3}\Bigr)\cos\chi\notag
\label{eq:Bfield}
\end{align}
In the equations above, $\chi$ is the inclination angle between the magnetic and rotational axes, while $b_\phi$ is the magnetic field generated through the 
shear of the dipolar field lines by the material in the disk. Among the different WG87 prescriptions for the poloidal 
field, we consider here the case of Eq.~\ref{eq:alf3} (the magnetospheric radius in all WG87 and W95 
prescriptions behave in a qualitatively similar way, see B09). According to this prescription, the value of  
$b_\phi$ at the upper ($z=h$) and inner ($R=R_{\rm M}$) surfaces of the disk are:   
\begin{align}
b^{(\rm upper)}_\phi&=-\Gamma\Bigl[1-\Bigl(\frac{\Omega_{\rm NS}}{\Omega_{\rm K}}\Bigr)\Bigr]B_z \\
b^{(\rm inner)}_\phi&=\Gamma\Bigl[1-\Bigl(\frac{\Omega_{\rm NS}}{\Omega_{\rm K}}\Bigr)\Bigr]B_R\notag, 
\label{eq:bipiccolo}
\end{align}
respectively ($\Gamma$=$\gamma$/$\alpha$). 
Assuming the case of a S73 thin disk, we also have $b_\phi(z)=-b_\phi(-z)$ and thus 
Eq.~\ref{eq:rmdef} becomes: 
\begin{equation}
\dot{M}\frac{d}{dR}\bigl(\Omega_{\rm K} r^2\bigr) = -h\frac{d}{dR}\bigl(R^2\langle b^{(\rm inner)}_\phi B_R\rangle\bigr)-b^{(\rm upper)}_\phi B_z R^2, 
\label{eq:sei}
\end{equation}
already calculated at $R$=$R_{\rm M}$. From this equation \citet{wang97} has obtained an expression for the magnetospheric radius\footnote{Note that in  
this treatment we are not taking into account the additional complication of possible vertical torques that might lead to the 
presence of a tilted accretion disk \citep[see, e.g.,][and references therein]{lai99}.}: 
\begin{align}
& \dot{M} \sqrt{GMR_{\rm M}} = \frac{2\Gamma\eta^2\mu^2}{R^3_{\rm M}} \cdot \notag \\
& \cdot \Bigl[(1-x_{\rm w}^{3/2})\cos^2\chi+\Bigl(\frac{h_0}{R_{\rm M}}\Bigr)(8-5x_{\rm w}^{3/2})\sin^2\chi\Bigr], 
\label{eq:seib}
\end{align}
where $h_0=h(R=R_{\rm M})$ is the disk height at $R_{\rm M}$ that can be obtained from S73 for the three different regions A, B, and C.  
The Eq.~\ref{eq:sei} and \ref{eq:seib} correspond to Eq.~6 and 7 of WG97. As neither the full derivation 
of these two equations, nor all required assumptions to obtain them were provided by WG97, we complete the current section 
with Appendix~\ref{appendix:1}. Note that Eq.~\ref{eq:seib} reduces to Eq.~\ref{eq:xdiff} for $\chi$=0~deg. 

\begin{figure}
\centering
\includegraphics[scale=0.58]{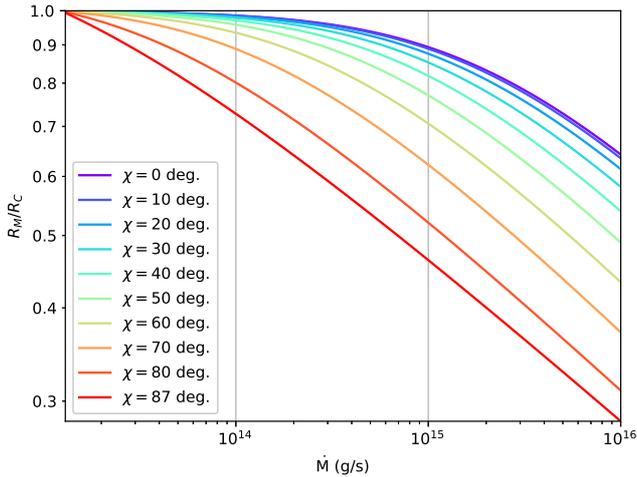}
\caption{Magnetospheric radius in units of the corotation radius as a function of the 
mass accretion rate obtained for the case of an inclined rotator from Eq.~\ref{eq:seib}. We assumed $P_{\rm spin}$=1.69~ms, 
$\alpha$=0.4, $\eta$=0.2, and $\gamma$=1.0. Different values for the inclination angle are shown in different colors for clarity. 
The magnetic moment corresponding to each angle is obtained from Eq.~\ref{eq:sp06} by assuming $\dot{P}_{\rm spin}$=5.39$\times$10$^{-21}$~s~s$^{-1}$.}     
\label{fig:plots2} 
\end{figure}

\begin{figure}
\centering
\includegraphics[scale=0.54]{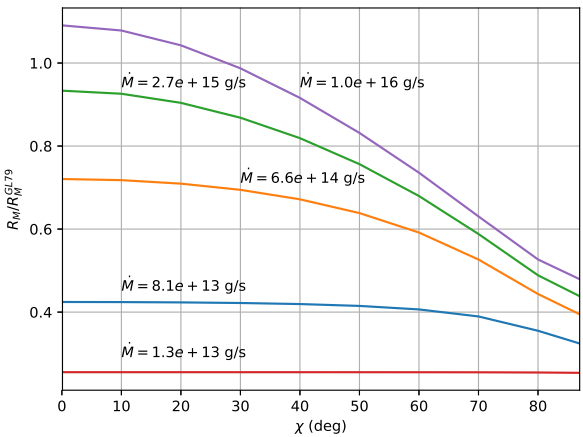}
\caption{Comparison between the magnetospheric radius calculated according to the WG97 (see Eq.~\ref{eq:seib}) and the 
GL79 (see Eq.~\ref{eq:rmGL}) prescriptions as a function of the inclination angle and for five values of the mass accretion rates 
(including the highest considered value of the mass accretion rate and the minimum value for which 
solutions to Eq.~\ref{eq:seib} are found for all considered inclination angles). 
We assumed $P_{\rm spin}$=1.69~ms, $\alpha$=0.4, $\eta$=0.2, and $\gamma$=1.0. The magnetic moment corresponding to each angle is 
obtained from Eq.~\ref{eq:sp06} by assuming $\dot{P}_{\rm spin}$=5.39$\times$10$^{-21}$~s~s$^{-1}$ and $P_{\rm spin}$=1.69~ms.}     
\label{fig:plots3} 
\end{figure}

The solutions to the full Eq.~\ref{eq:seib} were not reported by WG97. This author only showed the approximate decrease  
of the magnetospheric radius at high inclination angles using a simplified version of Eq.~\ref{eq:seib} where: (i) the radial dependence of the 
terms $|b^{(\rm upper)}/B_{\rm z}|$ and $|b^{(\rm inner)}/B_{\rm r}|$ is neglected and they are kept constant at a fixed value calculated at the magnetospheric radius; (ii)  
the term $h_0/R_{\rm M}$ is also assumed constant and fixed ($h_0/R_{\rm M}$=0.01), neglecting its dependence from $R_{\rm M}$ and the mass accretion rate.  
We show the solutions to the full Eq.~\ref{eq:seib} in Fig.~\ref{fig:plots2}, where we retained all functional dependences of the different terms. 
We assumed $\alpha$=0.4, $\eta$=0.2, $\gamma$=1.0, and a set of representative cases for the inclination angle. 
Compared to WG97, we also introduced a more self-consistent estimate of the NS magnetic moment which includes the dependence from the inclination angle, 
following the arguments by \citet{sp06}: 
\begin{equation}
\mu_{26}=2.6\times10^{11}(P_{\rm spin} \dot{P}_{\rm spin})^{1/2}(1+\sin^2(\chi))^{-1/2}~{\rm G~cm^3}.
\label{eq:sp06}
\end{equation}
Here $P_{\rm spin}$ is the NS spin period and $\dot{P}_{\rm spin}$ its derivative estimated from the radio pulsations. 
We used values representative of the fastest transitional millisecond pulsars $P_{\rm spin}$=1.69~ms and $\dot{P}_{\rm spin}$=5.39$\times$10$^{-21}$~s~s$^{-1}$ 
(see Sect.~\ref{sec:transitional}), such that $\mu_{26}$ ranges from 0.78 for $\chi$=0~deg to 0.55 for $\chi$=87~deg. 
In all cases, the magnetospheric radius is located within the region B of the S73 disk (see Eq.~\ref{eq:rab} and \ref{eq:bc}). 
In this region, the height of the disk practically scales linearly with the radius and thus $h_0/R_{\rm M}$ changes significantly 
with the mass accretion rate but not with the different inclination angles. For the specific set of parameters used in Fig.~\ref{fig:plots3}, 
$h_0/R_{\rm M}$$\simeq$0.002 at $\dot{M}_{16}\simeq$0.001 and $h_0/R_{\rm M}$$\simeq$0.007 at $\dot{M}_{16}\simeq$1. 

We note that solutions to the Eq.~\ref{eq:seib} at mass accretion rates lower than $\sim$10$^{13}$~g~s$^{-1}$ are only found for 
inclination angles lower than 50~deg (for the adopted set of the other parameters).   
In general, the minimum value of the mass accretion rate and maximum value of the inclination angle for which solutions to Eq.~\ref{eq:seib} 
exists depend strongly from the parameter $\eta$ (for a given parameter $\alpha$ and a NS with a given mass, radius, and magnetic moment). 
The larger is $\eta$, the larger (smaller) is the minimum accretion rate (maximum inclination angle) for which solutions exist.   
\begin{figure}
\centering
\includegraphics[scale=0.58]{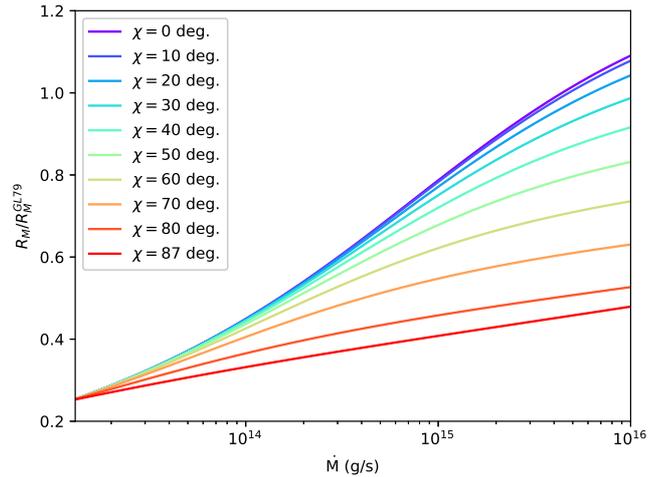}
\caption{Similar to Fig.~\ref{fig:plots3}, but here the comparison between the two magnetospheric radius prescriptions is shown 
as a function of the mass accretion rate and for different inclination angles.}     
\label{fig:plots4} 
\end{figure}

The interesting feature that emerges from Fig.~\ref{fig:plots2}  
is that the magnetospheric radius gets progressively smaller when larger inclination angles are considered (at comparable mass accretion rates). 
For $\chi$$\gtrsim$60~deg, the magnetospheric radius can be up to a factor of $\sim$2.5 smaller than that of the aligned rotator case presented by 
W87 and W95. The direct comparison between the magnetospheric radius 
of an inclined dipole according to the WG97 theory and the GL79 treatment is shown in Fig.~\ref{fig:plots3} and Fig.~\ref{fig:plots4} 
(note that Eq.~\ref{eq:rmGL} has been calculated in both figures by using the case $\chi=0$ for the NS magnetic moment 
in Eq.~\ref{eq:sp06}, as the GL79 approach is formally derived in case of an orthogonal rotator).   
Figure~\ref{fig:plots3} shows the ratio between the magnetospheric radius of W97 and GL79 as a function of the inclination 
angle for three values of the mass accretion rate. This is similar to Fig.~2 in WG97 but we made use of a full (and not simplified) 
solution to Eq.~\ref{eq:seib} and use $R_{\rm M}^{\rm GL79}$ instead of $R_{\rm A}$ in the comparison. 
Figure~\ref{fig:plots4} shows the ratio between the magnetospheric radius of W97 and GL79 as a function of the mass accretion rate for different inclination 
angles, highlighting the fact that the largest difference between the magnetospheric radius in the two treatments occurs toward lower mass 
accretion rates and higher inclination angles (with the effect of the inclination angle being more effective toward higher mass accretion rates).

\section{Application to low level accretion onto transitional millisecond pulsars}
\label{sec:transitional} 

Transitional millisecond pulsars are a sub-class of low mass X-ray binaries (LMXBs) hosting a NS which have been observed to 
switch between the rotation-powered to accretion-powered regimes \citep[see, e.g.][]{pulsar, tauris15}. So far, 3 confirmed systems 
have been identified: PSR\,J1023+0038 \citep{archibald09,patruno14}, XSS\,J12270-4859 \citep{saitou09,demartino10,demartino14}, 
and IGR\,J18245-2452 \citep{papitto13,ferrigno14}. The pulse period of the first two 
systems is strikingly similar, $\sim$1.69\,ms. IGR\,J18245-2452 hosts a NS spinning at 3.9\,ms. Two candidate systems, 3FGL\,J1544.6-1125 and 3FGL\,J0427.9-6704,  
have been suggested by \citet{bogdanov15} and \citet{strader16}, but no spin period has yet been reported for these sources.  
In the rotation-powered regime, the pressure of the NS dipole radiation is believed to push away the surrounding accretion disk and 
the compact object shines as a millisecond radio pulsar. In this state, only a moderate X-ray luminosity of 
$\lesssim$10$^{32}$~erg~s$^{-1}$ is recorded. This is ascribed to the presence of an intra-binary shock formed by the 
interaction between the pulsar wind and the material lost by the companion star \citep[see, e.g.,][]{bogdanov14,bassa14}. 
When the accretion disk is formed around the NS, the system switches to an accretion powered regime. 
So far, only IGR\,J18245-2452 displayed an accretion powered X-ray regime with a peak luminosity (10$^{36}$-10$^{37}$~erg~s$^{-1}$) and 
spectral/timing properties similar to those of classical accreting millisecond X-ray pulsars in outburst 
\citep[hereafter, AMXPs;][]{patruno12}. 
The two other confirmed systems and the candidate transitional pulsar likely underwent only some lower level accretion episodes, with typical 
luminosities of a few 10$^{33}$~erg~s$^{-1}$. This ``sub-luminous disk state'' \citep{linares14,papitto15} is usually characterized 
by a prominent variability in X-rays, and the presence of three distinct emission modes: a {\it low mode}, during which the luminosity 
can be as low as $\sim$5$\times$10$^{32}$~erg~$s^{-1}$, a {\it high mode} in which the typical luminosity is 
$\sim$(3-5)$\times$10$^{33}$~erg~$s^{-1}$, and a {\it flaring mode} where the luminosity can increase by another factor of $\sim$10. 
The switch between the different modes can be as fast as a few seconds, and 
the mechanism regulating these switches has not yet been understood \citep{bogdanov14}. 
During the high mode, the X-ray emission of transitional pulsars displays little variability and this is the only mode 
where X-ray pulsations could be clearly identified from both XSS\,J12270-4859 and PSR\,J1023+0038. 
For the first source, \citet{papitto15} measured a high mode 
average X-ray luminosity of $L_{\rm X}$$\sim$5$\times$10$^{33}$~erg~s$^{-1}$ (assuming a source distance of 1.40\,kpc) 
and a pulsed fraction of 6-7\% in the 0.5-11~keV energy range. For PSR\,J1023+0038, 
\citet{archibald14} measured an average high mode luminosity of $L_{\rm X}$$\sim$3$\times$10$^{33}$~erg~s$^{-1}$ (0.3-10~keV) and a pulsed fraction 
of about 8\% in a similar energy band (for a source distance of 1.37\,kpc). In both cases, 
the authors interpreted the observed X-ray pulsations as due to accretion onto the NS 
\citep[see also][who found indications for a rotationally-powered activity from PSR\,J1023+0038]{ambrosino17}.  

Although the phenomenologically complex sub-luminous disk state of the transitional pulsars is still lacking an 
exhaustive interpretation and many questions remain open \citep{campana16}, the discovery of accretion powered X-ray pulsations 
at the low X-ray luminosities of the high mode is particularly puzzling because it challenges the ``standard'' accretion scenario that is usually 
invoked to interpret the observations of transitional pulsars and other classes of accreting millisecond NS X-ray binaries (see below). 
In the following, we focus on this peculiar property of the high emission mode and in particular 
on the consideration of how low level accretion in these sources could still give rise to X-ray pulsations at such unprecedentedly 
low X-ray luminosity level. Our considerations are presently unable to explain also other phenomena observed in the sub-luminous disk 
state. We plan to discuss these in a future extension of this work. 

It is generally believed that accretion in a disk-fed NS LMXB can only take place 
as long as the NS magnetospheric radius is smaller than the corotation radius \citep[see, e.g.,][]{frank02}. 
According to this standard scenario, when $R_{\rm M}$ becomes larger than $R_{\rm c}$, it is expected that the 
centrifugal force at the boundary between the NS magnetosphere and material in the disk pushes the inflowing matter away, 
inhibiting accretion and driving outflows \citep[the so-called ``propeller effect'';][]{illarionov75}. To estimate the X-ray luminosity 
at which the onset of the propeller is expected in the case of the transitional pulsars, \citet{papitto15} and \citet{archibald14} used 
an expression for the magnetospheric radius very similar to the GL79 prescription (see Eq.~\ref{eq:rmGL}). 
As a result of this calculation they found that in both XSS\,J12270-4859 and PSR\,J1023+0038 accretion should be strongly inhibited 
already at luminosities $L_{\rm X}$$\gtrsim$2$\times$10$^{35}$~erg~s$^{-1}$. These authors thus suggested that the detection of X-ray pulsations in the sub-luminous 
disk state of these systems should not occur, unless very powerful outflows are generated by the rotating NS magnetosphere 
which remove $\gtrsim$95-99\% of the matter inflowing at the inner disk boundary. 
If this were the case, the large mass inflow rate would maintain 
$R_{\rm M}$$<$$R_{\rm c}$, such that low level accretion could take place in the standard scenario.  

While the presence of outflows in transitional pulsars has been supported by radio observations \citep[see, e.g.,][]{hill11,ferrigno14},  
a quantitative estimate of the ratio between the inflowing and outflowing mass rate from these systems cannot be reliably constrained 
yet from the observations. Advanced magnetohydrodynamic simulations of a rapidly rotating NS surrounded by a disk  
show that outflows can be generated in the so-called ``weak propeller regime'', when $R_{\rm M}$$\gtrsim$$R_{\rm c}$, but 
the mass ejection rate is at the most comparable to the mass accretion rate \citep[not larger than $\sim$20\% of the total mass inflow rate 
in the simulations of][]{ustyugova06}. In the strong propeller regime, when $R_{\rm M}$$\gg$$R_{\rm c}$, the ejection efficiency can reach about 
80\% \citep{lii14,lovelace13} that is still significantly lower than the level required in the cases 
of XSS\,J12270-4859 and PSR\,J1023+0038. In the strong propeller regime it is also 
unlikely that the bulk of the X-ray emission is dominated by residual accretion. As an example, in the case of 
IGR\,J18245-2452, the strong propeller regime has been invoked to explain the dramatic hardening of the source X-ray emission 
in terms of shocks that form between the outflows and the surrounding medium \citep[rather than by residual accretion; see, e.g.,][]{ferrigno14}. 

Only in the white dwarf binary AE Aquarii, evidence has been found for propelling efficiencies 
as high as 97\% \citep{oruru12}. Therefore, it has been suggested that such extreme values cannot be ruled out. 
We show below that the need of extreme ejection efficiencies to explain the 
X-ray pulsations of transitional pulsars in the sub-luminous disk state might not be needed if the WG97  
prescription of the magnetospheric radius is used in place of the GL79 simplified model. 

The luminosities at which pulsations have been detected in the X-ray emission of XSS\,J12270-4859 and PSR\,J1023+0038 
correspond to mass accretion rates of $\dot{M}$$\approx$$L_{\rm X}$$R_{\rm NS}$/(G$M_{\rm NS}$)=(2-3)$\times$10$^{13}$~g~s$^{-1}$. The 
approximate model of GL79 would give a magnetospheric radius at these very low mass accretion rates that is $\sim$3 times larger than the 
corotation radius, thus requiring the extreme outflows invoked by \citet{papitto15} and \citet{archibald14} 
to allow for (at least) some residual accretion and X-ray pulsations. 
According to WG97 calculations, the magnetospheric radius is still significantly smaller than the corotation radius 
even at mass accretion rates as low as $\sim$2-3$\times$10$^{13}$~g~s$^{-1}$. In case the NS is endowed with a large inclination angle ($\chi$$\gtrsim$60-70~deg), 
the magnetospheric radius is further reduced compared to the GL79 approach (even though this parameter is more effective at reducing the magnetospheric radius 
toward higher mass accretion rates; see Fig.~\ref{fig:plots2}). 
Under these assumptions, little to no outflows would thus be required to explain the pulsations at the very low X-ray luminosities recorded from 
XSS\,J12270-4859 and PSR\,J1023+0038 in the high mode of the sub-luminous-disk state. 

For all computations in Fig.~\ref{fig:plots2} we have assumed that the inclination angle also determines 
the effective dipole magnetic moment of the NS estimated through Eq.~\ref{eq:sp06} and a spin period derivative of 5.39$\times$10$^{-21}$~s~s$^{-1}$, as observationally 
measured in the case of PSR\,J1023+0038 \citep{archibald13}. These results are thus equally applicable to the case of XSS\,J12270-4859, as the spin period 
of the NS hosted in this system is virtually identical to that of PSR\,J1023+0038 and also the spin period derivatives of the two systems are fairly  
similar \citep{roy15}. 

Interestingly, some evidence for a large inclination angle ($>$60~deg) between the rotation and magnetic axis of the NS in XSS\,J12270-4859 
was provided by \citet{papitto15}, using also the results published by \citet{demartino14}. This makes our application of the W97 prescription for  
the magnetospheric radius calculation at large inclination angles promising for transitional millisecond pulsars in general.

\section{Discussion and conclusions}
\label{sec:conclusions} 

In this paper we reviewed the basic assumptions of the magnetically threaded disk model for accreting NS in X-ray binaries 
in both the original treatment presented by GL79 and the 
revised approach by W87 and WG95. The models make different predictions for the magnetospheric radius as a function of the mass accretion 
rate, with WG97 also extending the calculations to the case of an inclined dipole. 
The simplified approach of GL79 predicts that the magnetospheric radius is proportional to $\dot{M}^{-2/7}$.  
In the approach of W87 and W95 for an aligned rotator, the increase of the magnetospheric 
radius for decreasing mass accretion rates is slower and more complex.  
The most noticeable difference between these approaches occur at lower mass accretion rates.  
Moreover, when the full equations given in WG97 are solved, the magnetospheric radius is found to be further reduced at large 
inclination angles between the NS rotation and magnetic axes (with an effect more pronounced toward larger mass accretion rates). 
Assuming for consistency in all cases the same value of the screening parameter $\eta$=0.2, the magnetospheric radius obtained from WG97 
can be as small as $\sim$0.3 times the value expected from the GL79 calculations
either for low mass accretion rates ($\sim$10$^{13}$-10$^{15}$~g~s$^{-1}$) or for higher mass accretion rates ($\gtrsim$10$^{15}$~g~s$^{-1}$) 
and large inclination angles ($\chi\gtrsim60$~deg). 

We applied the magnetospheric radius prescription of W97 to the case of transitional millisecond pulsars, which are a sub-class of NS LMXBs 
showing a peculiar X-ray variability during their so-called sub-luminosity accretion state. The phenomenology observed during this state is 
complex, with three different emission modes identified (high, low, flaring) and rapid switches (a few seconds) 
occurring between them. So far there is still not an agreed scenario to explain all these behaviors, and we focused here in particular on the 
puzzling accretion-driven X-ray pulsations observed only during the high mode at an unprecedentedly low luminosity level (2-3$\times$10$^{33}$~erg~s$^{-1}$) 
compared to that of other previously known accreting millisecond X-ray pulsars in LMXBs ($\gtrsim$10$^{35}$~erg~s$^{-1}$). 
Following the usually adopted GL79 approach for the calculation of the magnetospheric radius, it is expected that at the mass accretion rates corresponding 
to a luminosity of 2-3$\times$10$^{33}$~erg~s$^{-1}$, the system should enter a very strong propeller regime with virtually no accretion taking 
place (and thus no detectable X-ray pulsations).   
Other works in the literature about the high mode of the transitional pulsars 
have proposed a scenario in which there is a substantial mass transfer from the companion to the NS  
to sufficiently compress the compact object magnetosphere and formally allow accretion, but at least $\sim$95-99\% of this material is  
ejected away by a very efficient propeller to explain the low luminosity at which pulsations are recorded \citep{papitto15, archibald14}. 
Even though these high propelling efficiencies can not be completely ruled out from the analogy with the white dwarf binary AE Aquarii, they are 
difficult to be reconciled with currently available MHD simulations of accreting NS in LMXBs. 
We showed that the revised magnetically threaded disk model presented by WG97 predicts a 
substantially smaller magnetospheric disk radius compared to GL79, especially when low mass accretion rates and high inclination angles between the 
NS magnetic and rotational axis are considered (assuming consistent values of the other involved parameters). 
If WG97 approach is used to estimate the 
magnetospheric radius, it is possible to envisage that accretion still takes place in the transitional pulsars when the 
mass accretion rate from the companion is as low as (2-3)$\times$10$^{-13}$~g~s$^{-1}$. This could potentially explain how to produce  
accretion-driven X-ray pulsations at a luminosity of 2-3$\times$10$^{33}$~erg~s$^{-1}$ without invoking very strong propellers.

\section*{Acknowledgments}
This publication was motivated by a team meeting
sponsored by the International Space Science Institute in Bern,
Switzerland. EB and AP thank ISSI for the financial
support during their stay in Bern. 
SA thanks the Department of Astronomy of the University of Geneva for the hospitality during part of this work.
AP acknowledges funding from the EUs Horizon 2020 Framework Programme for Research and Innovation under
the Marie Skodowska-Curie Individual Fellowship grant agreement 660657-TMSP-H2020-MSCA-IF-2014. 
LD acknowledges support by the Bundesministerium
f\"ur Wirtschaft und Technologie and the Deutsches
Zentrum f\"ur Luft und Raumfahrt through the grant FKZ 50 OG
1602. AP and LG acknowledges financial contribution from agreement ASI-INAF I/037/12/0 and 
ASI-INAF 2017-14-H.O. We thank the anonymous referee for the helpful comments. 
\bibliographystyle{aa}
\bibliography{swinging}

\appendix

\section{Full derivation of Eq.~20}
\label{appendix:1} 

We report here for reference the detailed calculations to derive Eq.~20. 
Following W97 approach, let us consider a NS endowed with a tilted dipolar magnetic field:
\begin{align}
B_R&=2\eta\Bigl(\frac{\mu}{R^3}\Bigr)\sin\chi\cos\phi\\
B_\phi&=\eta\Bigl(\frac{\mu}{R^3}\Bigr)\sin\chi\sin\phi+b_\phi\notag\\
B_z&=\eta\Bigl(\frac{\mu}{R^3}\Bigr)\cos\chi\notag. 
\label{eq:Bfield2}
\end{align}
Here, $\chi$ is the tilt angle between the NS magnetic and rotational axes, while $b_\phi$ is the magnetic field generated by the 
interaction of the NS magnetic field lines with the disk material in a Keplerian orbit around the compact object. 
This field is expressed at the upper ($z=h$) and inner ($R=R_{\rm 0}$) surface of the disk as follows: 
\begin{align}
b^{(\rm upper)}_\phi&=-\Gamma\Bigl[1-\Bigl(\frac{\Omega_{\rm NS}}{\Omega_{\rm K}}\Bigr)\Bigr]B_z \\
b^{(\rm inner)}_\phi&=\Gamma\Bigl[1-\Bigl(\frac{\Omega_{\rm NS}}{\Omega_{\rm K}}\Bigr)\Bigr]B_R\notag
\label{eq:bipiccolo}
\end{align}
Due to the symmetry of the problem $b_\phi(z)=-b_\phi(-z)$. We derive first Eq.~19 by starting from the Euler equation in a stationary state:
\begin{equation}
\rho(\vec{v}\cdot\nabla)\vec{v}=-\nabla P -\rho \nabla \Phi +\frac{1}{4\pi}\bigl(\nabla \times \vec{B}\bigr)\times\vec{B}
\label{eq:euler}
\end{equation}
According to Eq.~\ref{eq:Bfield2}, $B_z$ does not depend on $\phi$, thus the azimuthal 
component of the Lorentz force can be written as:
\begin{equation*}
\frac{1}{4\pi}\bigl[\bigl(\nabla \times \vec{B}\bigr)\times\vec{B}\bigr]_\phi=\frac{1}{4\pi}\Bigl[\frac{1}{R}B_\phi B_R+B_R\partial_R(B_\phi)+B_z\partial_z(B_\phi)-\frac{1}{R}B_R\partial_\phi (B_R)\Bigr]
\end{equation*}
Using the identity 
\begin{equation*}
\bigl(\vec{v}\cdot\nabla\bigr)\vec{v}=\frac{1}{2}\nabla v^2 + \bigl(\nabla \times \vec{v}\bigr)\times \vec{v}
\end{equation*}
we can rewrite the $\phi$ component of the left hand side of Eq.~\ref{eq:euler} as: 
\begin{equation*}
\begin{split}
&\rho\bigl[\frac{1}{2}\nabla v^2 + \bigl(\nabla \times \vec{v}\bigr)\times \vec{v}\bigr]_\phi=\rho\Bigl[\frac{1}{2R}\partial_\phi(v^2)+\frac{1}{R}v_\phi v_R+v_R\partial_R(v_\phi)+\\
&+v_z\partial_z(v_z)-\frac{1}{R}v_r\partial_\phi(v_r)-\frac{1}{R}v_z\partial_\phi(v_z)\Bigr]=\\
&=\frac{\rho}{R}v_R\partial_R(Rv_\phi)+\frac{\rho}{2R}\partial_\phi(v_\phi^2). 
\end{split}
\end{equation*}
Here we also assumed that the matter leaves the disk only once inside the magnetospheric radius and that for a thin disk $v_{\rm z}=0$.  
We thus obtain: 
\begin{equation}
\begin{split}
&\rho R v_R\partial_R(Rv_\phi)=\\ 
& \frac{1}{4\pi}\Bigl[B_\phi B_RR+R^2B_R\partial_R( B_\phi)+R^2B_z\partial_z (B_\phi)-RB_R\partial_\phi(B_R)\Bigr]+\\
& -\rho R\partial_\phi\Phi -R\partial_\phi(P)-\frac{\rho}{2}R\partial_\phi(v^2_\phi)
\end{split}
\label{eq:eulerintermedio1}
\end{equation}
Here the term $\partial_\phi \Phi$ vanishes due to the axial symmetry of gravitational potential.
Under the further assumption that $P$ and $v_{\phi}$ are $\phi$-independent\footnote{Note that the term 
$\partial_\phi(P)$ would be simplified later even in case $P$ is not assumed to be independent from $\phi$, as its integration 
between $0$ and $2\pi$ to be performed in the next steps fulfil the conditions expressed by Eq.~\ref{eq:null}.} 
with $v_{\phi} \simeq \Omega_K R$, we can write: 
\begin{equation}
\begin{split}
& \rho R v_R\partial_R(\Omega_{\rm K} R^2)= \\
& \frac{1}{4\pi}\Bigl[B_\phi B_RR+R^2B_R\partial_R( B_\phi)+R^2B_z\partial_z (B_\phi)-RB_R\partial_\phi(B_R)\Bigr]
\label{eq:eulerointermedio2}
\end{split}
\end{equation}
Equation~\ref{eq:eulerointermedio2} can be written in the following way: 
\begin{equation}
\rho R v_R\partial_R(\Omega_{\rm K}R^2)=\frac{1}{4\pi}\Bigl[\partial_R(R^2b_\phi B_R)+\partial_z(b_\phi B_z)R^2\Bigr]+A(R,\phi,z)
\label{eq:eulerointermedio3}
\end{equation}
where 
\begin{equation}
\int^{2\pi}_0A(R,\phi,z)d\phi=0. 
\label{eq:integrala}
\end{equation}
To demonstrate Eq.~\ref{eq:eulerointermedio3}, note that: 
\begin{equation*}
R^2B_R\partial_RB_\phi=\partial(R^2B_RB_\phi)-2RB_\phi B_R-R^2B_\phi\partial_RB_R,  
\end{equation*}
and 
\begin{equation*}
R^2B_z\partial_zB_\phi=R^2\partial_z(B_zB_\phi)-R^2B_\phi\partial_zB_z.  
\end{equation*}
We thus obtain for the right hand side of Eq.~\ref{eq:eulerointermedio2}
\begin{equation*}
\begin{split}
&=\frac{1}{4\pi}\bigl\{\partial_R(R^2B_RB_\phi)+R^2\partial_z(B_zB_\phi)-R^2B_\phi\bigl[\partial_R(B_R)+\partial_z(B_z)\bigr]+\\
&-RB_R\partial_\phi(B_R)-RB_\phi B_R\bigr\}
\end{split}
\end{equation*}
From the second Maxwell equation, we have: 
\begin{equation*}
\partial_R(B_R)+\partial_z(B_z)=-\frac{B_R}{R}-\frac{1}{R}\partial_\phi B_\phi, 
\end{equation*}
and thus: 
\begin{equation*}
\begin{split}
&\frac{1}{4\pi}\bigl[\partial_R(R^2B_RB_\phi)+R^2\partial_z(B_\phi B_z)+R^2B_\phi\bigl(\frac{B_R}{R}+\frac{1}{R}\partial_\phi(B_\phi)\bigr)\\
&-RB_R\partial_\phi(B_R)-RB_\phi B_z\bigr]=\\
&=\frac{1}{4\pi}\bigl[\underbrace{\partial_R(R^2B_RB_\phi)}_i+\underbrace{R^2\partial_z(B_\phi B_z)}_{ii}+\underbrace{RB_\phi\partial_\phi (B_\phi)}_{iii}-\underbrace{RB_R\partial_\phi(B_R)}_{iv}\bigr]
\label{eq:bific}
\end{split}
\end{equation*}
By using the relation 
\begin{equation}
B_\phi=b_\phi-\frac{1}{2}\partial_\phi(B_R)
\label{eq:bifi}
\end{equation}
that can be obtained from Eq.~\ref{eq:Bfield2}, we can rewrite the four terms above as:  
\begin{equation*}
\begin{split}
&i\rightarrow\partial_R(R^2B_RB_\phi)=\partial_R(R^2b_\phi B_R)-\frac{1}{2}\partial_R[R^2B_R\partial_\phi(B_R)]\\
&ii\rightarrow R^2\partial_z(B_\phi B_z)=R^2\partial_z(b_\phi B_z)-\frac{R^2}{2}\partial_z[B_z\partial_\phi (B_R)]\\
&iii\rightarrow RB_\phi\partial_\phi(B_\phi)= \frac{1}{2}R\partial_\phi(B^2_\phi)\\
&iv\rightarrow -RB_R\partial_\phi(B_R)= -\frac{1}{2}R\partial_\phi(B^2_R),\\
\end{split}
\end{equation*}
and thus: 
\begin{equation*}
\begin{split}
&A(R,\phi,z)=\\
& -\frac{1}{8\pi}\Bigl\{\underbrace{\partial_R\bigl[R^2B_R\partial_\phi(B_R)\bigr]}_\alpha+\underbrace{R^2\partial_z\bigl[B_z\partial_\phi(B_R)\bigr]}_\beta
+\underbrace{R\partial_\phi(B^2_R-B^2_\phi)}_\gamma\Bigr\}\\
\end{split}
\end{equation*}
To prove Eq.~\ref{eq:integrala}, it is sufficient to note that $B_z$ does not depend on $\phi$ and that for a generic function $f(R, \phi, z)$:
\begin{equation}
\int^{2\pi}_0\partial_\phi( f )d\phi=f(2\pi)-f(0)=0
\label{eq:null}
\end{equation} 

Equation~\ref{eq:eulerointermedio3} has now to be integrated over $\phi$ from $0$ to $2\pi$ and over $z$ from $-h$ to $h$. 
Taking into account the simplifications possible under the assumption of a thin disk case and the continuity equation: 
\begin{equation*}
\dot{M}=-\int^{2\pi}_0Rd\phi\int^h_{-h}\rho v_R dz,  
\end{equation*}
we obtain for the left hand side of Eq.~\ref{eq:eulerointermedio3}: 
\begin{equation*}
\begin{split}
&\int^{2\pi}_0d\phi\int^h_{-h}dz\rho R v_R\partial_R(\Omega_{\rm K}R^2)=\frac{d}{dR}(\Omega_{\rm K}R^2)R\int^{2\pi}_0d\phi\int^h_{-h}dz\rho v_R=\\
&=-\dot{M}\frac{d}{dR}(\Omega_{\rm K}R^2)
\end{split}
\end{equation*}
For the first and second right hand side term of Eq.~\ref{eq:eulerointermedio3}, we have: 
\begin{equation*}
\begin{split}
&\int^{2\pi}_0d\phi\int^h_{-h}dz\partial_R\Bigl(R^2\frac{b_\phi B_R}{4\pi}\Bigr)=\frac{1}{4\pi}\int^{2\pi}_0d\phi\partial_R\Bigl(R^2\int^h_{-h}b_\phi B_Rdz\Bigr)=\\
&=\frac{h}{2\pi}\frac{d}{dR}\Bigl(R^2\int^{2\pi}_0b^{(\rm inner)}_\phi B_Rd\phi\Bigr)=h\frac{d}{dR}\bigl(R^2\langle b^{(\rm inner)}_\phi B_R\rangle\bigr)
\end{split}
\end{equation*}
and 
\begin{equation*}
\int^{2\phi}_0d\phi\int^h_{-h}dzR^2\partial_z\Bigl(\frac{b_\phi B_z}{4\pi}\Bigr)=\frac{R^2}{4\pi}\int^{2\pi}_0d\phi \bigl[b_\phi B_z\bigr]^h_{-h}=R^2b^{(\rm upper)}_\phi B_z, 
\end{equation*}
respectively. 
Putting all terms together, we get at $R$=$R_{\rm M}$: 
\begin{align*}
\dot{M}\frac{d}{dR}\bigl(\Omega_{\rm K}r^2\bigr)=\underbrace{-h\frac{d}{dR}\bigl(R^2\langle b^{(\rm inner)}_\phi B_R\rangle\bigr)}_{I}-\underbrace{b^{(\rm upper)}_\phi B_z R^2}_{II} 
\end{align*}
For the left hand side, we have: 
\begin{equation*}
\dot{M}\frac{d}{dR}\bigl(\Omega_{\rm K}r^2\bigr)_{R_{\rm M}}=\dot{M}\frac{d}{dR}\bigl(\sqrt{GMR}\bigr)_{R_{\rm M}}=\frac{\dot{M}GM}{2\sqrt{R_{\rm M}}}
\end{equation*}
For the two terms on the right hand side, we have:
\begin{equation*}
\begin{split}
I \rightarrow &-h\frac{d}{dR}(R^2\langle b^{(\rm inner)}_\phi B_R\rangle)=\\
&-\frac{h\Gamma}{2\pi}\frac{d}{dR}\Bigl\{R^2\Bigl[1-\Bigl(\frac{\Omega_{\rm NS}}{\Omega_{\rm K}}\Bigr)\Bigr]\int^{2\pi}_0B^2_Rd\phi\Bigr\}=\\
&=\frac{h\Gamma\eta^2\mu^2}{R^5}\sin^2\chi\Bigl[8-5\Bigl(\frac{\Omega_{\rm NS}}{\Omega_{\rm K}}\Bigr)\Bigr]\rightarrow\frac{h_0\Gamma\eta^2\mu^2}{R^5_0}\sin^2\chi(8-5\omega)
\end{split}
\end{equation*}
and 
\begin{equation*}
\begin{split}
II\rightarrow &-b^{(\rm upper)}_\phi B_zR^2=\Gamma R^2 \Bigl[1-\Bigl(\frac{\Omega_{\rm NS}}{\Omega_{\rm K}}\Bigr)\Bigr]B^2_z= \\ 
&\rightarrow\frac{\Gamma\eta^2\mu^2}{R^4}\cos^2\chi(1-\omega). 
\end{split}
\end{equation*}
We finally get: 
\begin{equation*}
\dot{M}\sqrt{GMR_{\rm M}}=\frac{2\Gamma\eta^2\mu^2}{R^3_0}\Bigl[(1-\omega)\cos^2\chi+\Bigl(\frac{h_0}{R_{\rm M}}\Bigr)(8-5\omega)\sin^2\chi\Bigr]. 
\end{equation*}

\end{document}